\documentclass[pra,twocolumn,amsmath,amssymb,superscriptaddress,showpacs]{revtex4}
\usepackage{graphics}
\usepackage{epsfig}
\usepackage{bbm}
\usepackage[latin1]{inputenc} 

\newcommand{\be}{\begin{equation}}
\newcommand{\ee}{\end{equation}}
\newcommand{\eea}{\end{eqnarray}}
\newcommand{\bea}{\begin{eqnarray}}
\newcommand{\exs}[1]{\ensuremath{\langle{#1}\rangle}}
\newcommand{\eins}{\ensuremath{\mathbbm 1}}
\newcommand{\qed}{\ensuremath{\hfill \Box}}

\newcommand{\CC}{\ensuremath{\mathcal{C}}}

\newcommand{\VV}{\ensuremath{\mathcal{V}}}

\newcommand{\BB}{\ensuremath{\mathcal{B}}}
\newcommand{\ketbra}[1]{\ensuremath{| #1 \rangle \langle #1 |}}
\newcommand{\ket}[1]{\ensuremath{|#1\rangle}}

\newcommand{\kommentar}[1]{}
\newcommand{\EQ}[1]{Eq.~(\ref{#1})}
\newcommand{\EE}{\ensuremath{\mathcal{E}}}
\newcommand{\LL}{\ensuremath{\mathcal{L}}}
\newcommand{\NN}[1]{\ensuremath{\mathcal{N}(#1)}}
\newcommand{\NNT}[1]{\ensuremath{\tilde{\mathcal{N}}(#1)}}
\newcommand{\mean}[1]{\ensuremath{\langle #1 \rangle}}

\begin{document}
\title{Two-setting Bell Inequalities for Graph States}
\date{\today}
\begin{abstract}
We present Bell inequalities for graph states with a high violation of local realism.
In particular, we show that there is a basic Bell
inequality for every nontrivial
graph state which is violated by the state at least by a
factor of two. This inequality needs the measurement of at most
two operators for each qubit and involves only some of the qubits.
We also show that for some families of graph states
composite Bell inequalities can be constructed such that
the violation of local realism increases
exponentially with the number of qubits.
We prove that some of our inequalities are facets of the convex
polytope containing the many-body correlations consistent
with local hidden variable models.
Our Bell inequalities are built from stabilizing operators of graph states.
\end{abstract}

\author{G\'eza T\'oth}
\email{toth@alumni.nd.edu}
\affiliation{Max-Planck-Institut f\"ur
Quantenoptik, Hans-Kopfermann-Stra{\ss}e 1, D-85748 Garching,
Germany,}

\affiliation{Research Institute of Solid State Physics and Optics, Hungarian Academy of Sciences, H-1525 Budapest P.O. Box 49, Hungary}

\author{Otfried G\"uhne}
\email{otfried.guehne@uibk.ac.at}
\affiliation{Institut f\"ur
Quantenoptik und Quanteninformation, \"Osterreichische Akademie
der Wissenschaften, A-6020 Innsbruck, Austria,}

\author{Hans J. Briegel}
\email{hans.briegel@uibk.ac.at}
\affiliation{Institut f\"ur Quantenoptik und Quanteninformation,
\"Osterreichische Akademie der Wissenschaften, A-6020 Innsbruck,
Austria,} \affiliation{Institut f\"ur Theoretische Physik,
Universit\"at Innsbruck, Technikerstra{\ss}e 25, A-6020 Innsbruck,
Austria}


\maketitle

\section{Introduction}

Bell inequalities \cite{B64,ZB02,WW01,mermin,ardehali,PR92,PS01,F81} have
already been used for several decades as an essential tool for
pointing out the impossibility of local realism in describing the
results arising from correlation measurements on quantum states.
While the relatively young theory of quantum entanglement
\cite{W89} is also used to characterize the
non-classical behavior of quantum
systems, Bell inequalities still remain essential both from the
fundamental point of view and also from the point of view of
quantum information processing applications. For example, the
violation of a two-setting Bell inequality indicates that there is
a partition of the multi-qubit quantum state to two parties such
that some pure entanglement can be distilled \cite{A02}.
Furthermore, any state which violates a Bell inequality can be
used for reducing the communication complexity of certain tasks
\cite{zukbruk}.

This paper is devoted to the study of the non-local properties of
graph states. Graph states \cite{graph1,graph2,graph3,
experiments} are a family of multi-qubit states which comprises
many useful quantum states such as the Greenberger-Horne-Zeilinger
(GHZ, \cite{GH90}) states and the cluster states \cite{cluster}.
They play an important role in applications: Measurement-based
quantum computation uses graph states as resources
\cite{graphapp1th,graphapp1ex} and all codewords in the standard
quantum error correcting codes correspond to graph states
\cite{graphapp2}.

From a theoretical point of view one remarkable fact about graph
states is that they can elegantly be described in terms of their
{\it stabilizing operators.} This means that a graph state can be
defined as an eigenstate of several such locally measurable
observables. These observables form a commutative group called
{\it stabilizer} \cite{G96}. Stabilizer theory has already been
used to study the nonlocal properties of special instances of
graph states \cite{DP97,S04}. In a previous work we showed that
for every non-trivial graph state it is possible to construct
three-setting Bell inequalities which are maximally violated only
by this state \cite{GTHB04}. These inequalities use all the
elements of the stabilizer.

In this paper we will examine how to create efficient two-setting
Bell inequalities for graph states by using only some of the
elements of the stabilizer. Efficiency in this case means that our
inequalities allow for a high violation of local realism. Apart
from trivial graphs, this is at least a factor of two and
increases exponentially with the size for some families of  graph
states. Interestingly, our inequalities are Mermin- and
Ardehali-type inequalities with multi-qubit observables
\cite{mermin,ardehali}. Our Mermin-type inequalities are based on
a GHZ-type violation of local realism \cite{S04}. This means that
all correlation terms in our Bell inequalities are $+1$ for a
given graph state, while there is not a local hidden variable
(LHV) model with such correlations.

Our paper is organized as follows. In Sec.~II we recall the basic
facts about graph states and the stabilizer formalism. We also
explain the notation we use to formulate our Bell inequalities and
provide a first example for a Bell inequality for graph states. In
Sec.~III we present Bell inequalities for general graphs involving
the stabilizing operators of a vertex and its neighbors. These are
Mermin-type inequalities with multi-qubit observables. Then, in
Sec.~IV we discuss how to construct inequalities having a
violation of local realism increasing exponentially with the
number of qubits for some families of graph states. In Sec.~V we
present Ardehali-type inequalities with multi-qubit observables
which have a higher violation in some cases than the Mermin-type
inequalities. In Sec.~VI we will show that some of our
inequalities correspond to facets (maximal faces) of the convex
polytope containing correlations allowed by LHV models. Finally,
in Sec.~VII we will discuss the connection of our inequalities to
existing inequalities for four-qubit cluster states.

\section{Definitions and notations}

Let us start by briefly recalling the definition of graph states.
A detailed investigation of graph states can be found in
Ref.~\cite{graph2}. A graph state corresponds to a graph $G$
consisting of $n$ vertices and some edges. Some simple graphs are
shown in Fig.~1. Let us characterize the connectivity of this
graph by $\NN{i}$, which gives the set of neighbors for vertex
$i.$ We can now define for each vertex a locally measurable
observable via
\begin{eqnarray}
g_k:=X^{(k)} \prod_{l\in \NN{k}} Z^{(l)}, \label{stab}
\end{eqnarray}
where $X^{(k)}$ and $Z^{(k)}$ are Pauli spin matrices.
A graph state $\ket{G_n}$ of $n$ qubits is now defined
to be the state which has these  $g_k$ as stabilizing
operators. This means that the $g_k$ have the state
$\ket{G_n}$ as an eigenstate with eigenvalue $+1,$
\begin{equation}
g_k\ket{G_n}=\ket{G_n}.
\end{equation}
In fact, not only the $g_k$, but also their products (for example
$g_1g_2$, $g_1g_3g_4$, etc.) stabilize the state $\ket{G_n}.$
These $2^n$ operators will be denoted by $\{S_m\}_{m=1}^{2^n}.$
They form a commutative group called the {\it stabilizer} and
$\{g_k\}_{k=1}^n$ are the generators of this group \cite{G96}. All
the elements $S_m$ of the stabilizer are products of Pauli spin
matrices.

\begin{figure}
\centerline{\epsfxsize=3in \epsffile{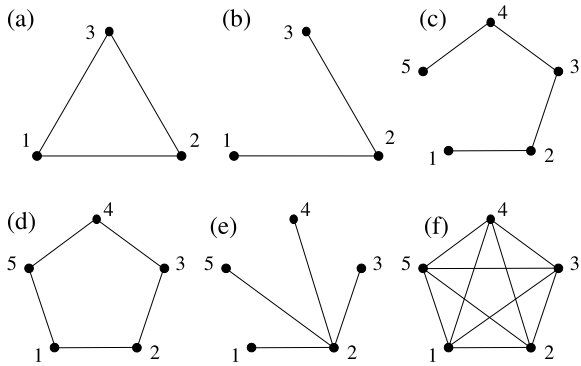}}
\caption{Examples of graphs: (a) The three-vertex fully connected
graph $FC_3.$ (b) The three-vertex linear cluster  graph $LC_3.$
(c) The five-vertex linear cluster graph $LC_5.$ (d) The ring
cluster  graph $RC_5.$ (e) The star graph $SC_5.$ (f) The fully
connected  graph $FC_3$ with five vertices. Examples of more
complicated graphs are shown in Fig.~2. } \label{bellgraf}
\end{figure}

Let us fix the notation for formulating Bell inequalities.
A Bell operator is typically presented as the sum of many-body
correlation terms. Now we will consider Bell operators $\BB$ which
are the sum of some of the stabilizing operators
\begin{equation}
\BB=\sum_{m\in J} S_m, \label{BB}
\end{equation}
where set $J$ tells us which stabilizing operators we use for
$\BB.$ Since all $S_m$ are the products of Pauli spin matrices
$X^{(k)}$, $Y^{(k)}$ and $Z^{(k)}$, we naturally assume that for
our Bell inequalities for each qubit these spin coordinates are
measured.

The maximum of the mean value $\mean{\BB}$ for quantum states can
immediately be obtained: For the graph state $\ket{G_n}$ all
stabilizing operators have an expectation value $+1$ thus
$\mean{\BB}$ is an integer and it equals the number of stabilizing
operators used for constructing $\BB$ as given in Eq.~(\ref{BB}).
Clearly, there is no quantum state for which $\mean{\BB}$ could be
larger.

Now we will determine the maximum of $\mean{\BB}$ for LHV models.
It can be obtained in the following way: We take the definition
Eq.~(\ref{BB}). We then replace the Pauli spin matrices with real
(classical) variables $X_k$, $Y_k$ and $Z_k.$ Let us denote the
expression obtained this way by $\EE(\BB)$:
\begin{equation}
\EE(\BB):=[\BB] \raisebox{-0.11em}{$\big\vert_{X^{(k)}\rightarrow X_k,Y^{(k)}\rightarrow Y_k,Z^{(k)}\rightarrow Z_k}$}.
\end{equation}
It is known that when maximizing $\EE(\BB)$ for LHV models it is
enough to consider deterministic LHV models which assign a
definite $+1$ or $-1$ to the variables $X_k$, $Y_k$ and $Z_k.$ The
value of our Bell operators for a given deterministic local model
$\LL,$ i.e., an assignment of  $+1$ or $-1$ to the classical
variables will be denoted by $\EE_{\LL}(\BB).$ Thus we can obtain
the maximum of the absolute value of $\mean{\BB}$ for LHV models
as
\begin{equation}
\CC(\BB):=\max_{\LL} |\EE_{\LL}(\BB)|.
\end{equation}

The usefulness of a Bell inequality in experiments can be
characterized by the ratio of the quantum and the classical
maximum
\begin{equation}
\VV(\BB):=\frac {\max_{\Psi} |\exs{\BB}_\Psi|} {\max_{\LL}
|\EE_{\LL}(\BB)|}.\label{VV}
\end{equation}
Thus $\VV(\BB)$ is the maximal violation of local realism allowed
by the Bell operator $\BB.$ For LHV models we have
$\mean{\BB}\le\CC(\BB).$ If $\VV(\BB)>1$ then this inequality is a
Bell inequality, and some quantum states violate it. In general,
the larger $\VV(\BB)$ is, the better our Bell inequality is.

As a warming up exercise, let us now write down explicitly a Bell
inequality for a three-qubit linear cluster state $\ket{LC_3}$
[see Fig.~1(b)]. We define
\begin{eqnarray}
\BB^{(LC_3)}&:= &Z^{(1)}X^{(2)}Z^{(3)} + Y^{(1)} Y^{(2)}Z^{(3)} +
\nonumber \\
&+&Z^{(1)} Y^{(2)}Y^{(3)} - Y^{(1)} X^{(2)}Y^{(3)}.\label{CCC3}
\end{eqnarray}
Here $\BB^{(LC_3)}$ is given as the sum of
four stabilizing operators of $\ket{LC_3}.$
For $\ket{LC_3}$ all of these terms have an expectation value $+1$
and the maximum of $\mean{\BB^{(LC_3)}}$ for quantum states is $4.$
For classical variables, we have
\begin{eqnarray}
Z_1X_2Z_3+Y_1Y_2Z_3+Z_1Y_2Y_3-Y_1X_2Y_3 \le 2,
\label{C3}
\end{eqnarray}
thus we have $\CC(\BB^{(LC_3)})=2.$ The maximal violation of local
realism is $\VV(\BB^{(LC_3)})=2.$ The inequality Eq.~(\ref{C3}),
apart from relabeling the variables, has been presented by Mermin
\cite{mermin} for GHZ states. This is not surprising, since the
state $\ket{LC_3}$ is, up to local unitary transformations, the
GHZ state \cite{graph2}. It is instructive to write down the Bell
operator of Eq.~(\ref{C3}) with the $g_k^{(LC_3)}$ operators of
the three-qubit linear cluster state
\begin{eqnarray}
\BB^{(LC_3)}&=&g_2^{(LC_3)}(1+g_3^{(LC_3)})(1+g_1^{(LC_3)}),
\label{B3}
\end{eqnarray}
where, based on Eq.~(\ref{stab}), we have
$g_1^{(LC_3)}=X^{(1)}Z^{(2)}$,
$g_2^{(LC_3)}=Z^{(1)}X^{(2)}Z^{(3)}$ and
$g_3^{(LC_3)}=Z^{(2)}X^{(3)}.$ Now the question is what happens if
new vertices are added to our three-vertex linear graph and spins
$1$, $2$ and $3$ have new neighbors as shown in Fig.
\ref{fig_graphs}(a). For this case the Bell operator of the form
Eq.~(\ref{B3}) will be generalized in the following.

Finally, before starting our main discussion, let us recall one
important fact which simplifies the calculation of the maximum
mean value for local realistic models:
\\
{\bf Lemma 1.} Let $\BB$ be a Bell operator consisting of a subset
of the stabilizer for some graph state. Then, when computing the
classical maximum $\CC(\BB)$ one can restrict the attention to LHV
models which assign $+1$ to all
$Z_k.$ \\
{\it Proof.} The proof of this fact was given in
Ref.~\cite{GTHB04}, we repeat it here for completeness. From the
construction of  graph states and the multiplication rules for
Pauli matrices it is easy to see that for an element $S$ of the
stabilizer the following fact holds: we have $Y^{(i)}, Z^{(i)}$ at
the qubit $i$ in $S$ iff the number of $Y^{(k)}$ and $X^{(k)}$ in
the neighborhood $\NN{i}$ in $S$ is odd. Thus, if a LHV model
assigns $-1$ to $Z^{(i)}$, we can, by changing the signs for
$Z^{(i)},Y^{(i)}$ and for all $X^{(k)}$ and all $Y^{(k)}$ with
$k\in \NN{i},$ obtain a LHV model with the same mean value of
$\BB$ and the desired property. $\qed$

\section{Bell inequality associated with a vertex and its neighborhood}

\begin{figure}
\centerline{\epsfxsize=3in \epsffile{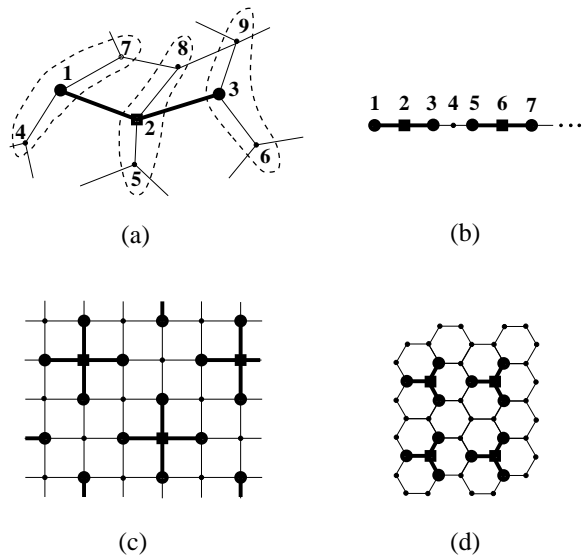} }
\caption{Bell inequalities for graph states. (a) The graphical
representation of a Bell inequality involving the generators of
vertices $1$, $2$ and $3$. The corresponding three-vertex subgraph
is shown in bold. The dashed lines indicate the three qubit-groups
involved in this inequality. For the interpretation of symbols at
the vertices see text. (b) Graphical representation for Bell
inequalities for linear cluster states, (c) a two-dimensional
lattice and (d) a hexagonal lattice. } \label{fig_graphs}
\end{figure}

Now we describe a  method which assigns a Bell inequality to each
vertex in the graph. The inequality is constructed such that it is
maximally violated by the state $\ket{G}$, having stabilizing
operators $g_k.$
\\
{\bf Theorem 1.} Let $i$ be a vertex and let $I \subseteq \NN{i}$
be a subset of its neighborhood, such that none of the vertices in
$I$ are connected by an edge. Then the following operator
\begin{eqnarray}
\BB(i,I):=g_i \prod_{j\in I} (1+g_j) \label{bbi_bell},
\end{eqnarray}
defines a Bell inequality $|\mean{\BB(i,I)}|\leq L_M(|I|+1)$ with
\begin{eqnarray}
L_M(m) := \left\{
\begin{array}{ll}
2^{\frac{m-1}{2}} & \textrm{ for odd $m$, }
\\
2^{\frac{m}{2}} & \textrm{ for even $m$, }
\end{array}
\right. \label{bbi_bell_cc}
\end{eqnarray}
and $\ket{G}$ maximally violates it with
$\exs{\BB(i,I)}=2^{|I|}$.
The notation $\BB(i,I)$ indicates that the Bell operator for our
inequality is constructed with the generators corresponding to
vertex $i$ and to some of its neighbors given by set $I$.
\\
{\it Proof.} Let us first consider a concrete example shown
in Fig. \ref{fig_graphs}(a). Consider vertex $2$ and its
two neighbors, vertices $1$ and $3$. Now constructing
the Bell operator $\BB$ involves $g_2$ and $(1+g_{1/3})$.
This is expressed by denoting these vertices by squares
and disks, respectively. The three-vertex subgraph with
bold edges now represents the Bell inequality $\BB(2,\{1,3\})$.
Then by expanding the brackets in \EQ{bbi_bell} one obtains
\bea
&&\BB(2,\{1,3\})=Z^{(1)}(Z^{(5)}Z^{(8)}X^{(2)})Z^{(3)}
\nonumber\\
&&\;\;\;\;+(Z^{(4)}Z^{(7)}Y^{(1)})(Z^{(5)}Z^{(8)}Y^{(2)})Z^{(3)}
\nonumber\\
&&\;\;\;\;+Z^{(1)}(Z^{(5)}Z^{(8)}Y^{(2)})(Z^{(6)}Z^{(9)}Y^{(3)})
\nonumber\\
&&\;\;\;\;-(Z^{(4)}Z^{(7)}Y^{(1)})(Z^{(5)}Z^{(8)}X^{(2)})(Z^{(6)}Z^{(9)}Y^{(3)}).
\nonumber\\
\label{bellvertex} \eea
Eq.~(\ref{bellvertex}) corresponds to  measuring two
multi-spin observables at each of three parties, where a
party is formed out of several qubits. In fact, in
Eq.~(\ref{bellvertex}) one can recognize the three-body
Mermin inequality with multi-qubit observables. These
observables are indicated by bracketing.
The corresponding three qubit groups are indicated by
dashed lines in Fig.~\ref{fig_graphs}(a).
Note that if vertices $1$ and $3$ were connected by an edge
then an operator of the form Eq.~(\ref{bbi_bell}) would involve
three different variables at sites $1$ and $3$.

Let us now turn to the general case of an inequality involving a
vertex and its $N_{{\rm neigh}}\ge 2$ neighbors given
in $\{I_k\}_{k=1}^{{\rm N_{neigh}}}.$ Then, similarly to the previous
tripartite example, this inequality is effectively a
$(|I|+1)$-body Mermin's inequality. In order to see that, let us
define the reduced neighborhood of vertex $k$ as
\begin{equation}
\NNT{k}:=\NN{k} \backslash (I\cup \{i\}).
\end{equation}
Then we define the following multi-qubit observables
\begin{eqnarray}
{A}^{(1)}&:=& Y^{(i)} \prod_{k\in \NNT{i}} Z^{(k)},
\nonumber\\
{B}^{(1)}&:=& X^{(i)} \prod_{k\in \NNT{i}} Z^{(k)},
\nonumber\\
{A}^{(j+1)}&:=& Z^{(I_{j})},\nonumber\\
{B}^{(j+1)}&:=& Y^{(I_{j})} \bigg( \prod_{k\in \NNT{I_{j}}}
Z^{(k)}\bigg), \label{nonlocalmermin}
\end{eqnarray}
for $j=1,2,...,N_{\rm{neigh}}$ and $I_j$ denotes the $j$-th element of $I.$
Then we can write down our Bell
operator given in Eq.~(\ref{bbi_bell}) as the Bell operator of a
Mermin inequality with $A^{(i)}$ and $B^{(i)}$
\begin{eqnarray}
\BB(i,I)
& = &
\sum_\pi B^{(1)}A^{(2)}A^{(3)}A^{(4)}A^{(5)}\cdot\cdot\cdot
\nonumber\\
&-&\sum_\pi B^{(1)}B^{(2)}B^{(3)}A^{(4)}A^{(5)}\cdot\cdot\cdot
\nonumber\\
&+&\sum_\pi B^{(1)}B^{(2)}B^{(3)}B^{(4)}B^{(5)}\cdot\cdot\cdot,
\nonumber\\
\label{BBB}
\end{eqnarray}
where $\sum_\pi$ represents the sum of all possible permutations
of the qubits that give distinct terms. Hence the bound for local
realism for Eq.~(\ref{bbi_bell}) is the same as for the
$(|I|+1)$-partite Mermin inequality \cite{mermin, MERMIN}. For
$\ket{G}$ all the terms in the Mermin inequality using the
variables defined in Eq.~(\ref{nonlocalmermin}) have an
expectation value $+1$ thus $\exs{\BB(i,I)}=2^{|I|}. \qed$

There is also an alternative way to understand why the extra
$Z^{(k)}$ terms in the Bell operator does not influence the
maximum for LHV models. I.e., the maximum is the same as for the
$(|I|+1)$-qubit Mermin inequality. For that we have to use Lemma~1
described for computing the maximum for LHV models for an
expression constructed as a sum of stabilizer elements of a graph
state. Lemma~1 says that the $Z_k$ terms can simply be set to $+1$
and for computing the maximum it is enough to vary the $X_k$ and
$Y_k$ terms. Thus from the point of view of the maximum, the extra
$Z_k$ terms can be neglected and would not change the maximum for
LHV models even if it were not possible to reduce our inequality
to a $(|I|+1)$-body Mermin inequality using the definitions
Eq.~(\ref{nonlocalmermin}).

Furthermore, it is worth noting that the above presented
inequalities can be viewed as conditional Mermin inequalities
for qubits $\{ i \} \cup I$ after $Z^{(j)}$ measurements
on the neighboring qubits
are performed  \cite{JIC}. Indeed, after measuring $Z^{(j)}$ on
these qubits, a state locally equivalent to a GHZ state is
obtained. Knowing the outcomes of the $Z^{(j)}$ measurements, one
can determine which state it is exactly and can write down a
Mermin-type inequality with two single-qubit measurements per site
which is maximally violated by this state. Indeed, this
Mermin-type inequality can be obtained from the Bell inequality
presented in Eqs.~(\ref{bbi_bell}-\ref{bbi_bell_cc}) in Theorem 1,
after substituting in it the $\pm1$ measurement results for these
$Z^{(j)}$ measurements. Our scheme shows some relation to the Bell
inequalities presented in Ref. \cite{PR92}. These were essentially
two-qubit Bell inequalities conditioned on measurement results on
the remaining qubits.

Finally, we can state: \\
 {\bf Theorem 2.} Every nontrivial graph
state violates a two-setting Bell inequality at least by a factor
of two.
\\
{\it Proof.} Every nontrivial graph state has at least one vertex
$i$ with at least two neighbors, $j$ and $k$. There are now two
possibilities: (i) If these two neighbors are not connected to
each other by an edge then Theorem 1 provides a Bell inequality
with a Bell operator $\BB(i,\{j,k\})$ which is violated by local
realism by a factor of two. (ii) If these two neighbors are
connected by an edge, then the situation of Fig.1(a) occurs. In
this case, we  look at the Bell operator \be
\BB^{(FC_3)}:=g_1^{(FC_3)}+g_2^{(FC_3)}+ g_3^{(FC_3)}+
g_1^{(FC_3)}g_2^{(FC_3)}g_3^{(FC_3)}. \ee As before, one can now
show that this results in a Bell inequality which is equivalent to
the three-qubit Mermin inequality. $\qed$

\section{Composite Bell inequalities}

Theorem 1 can also be used to obtain families of
Bell inequalities with a
degree of violation increasing exponentially with the number of
qubits. In order to do this, let us start from two Bell
inequalities of the form
\begin{eqnarray}
|\EE_1| \le \CC_1,  \nonumber\\
|\EE_2| \le \CC_2. \label{tb}
\end{eqnarray}
where $\EE_{1/2}$ denote two expressions with classical variables
$X_k$'s, $Y_k$'s and $Z_k$'s. Then it follows immediately that
\begin{equation}
|\EE_1 \EE_2 | \le \CC_1 \CC_2. \label{BBBB}
\end{equation}
Concerning Bell inequalities, one has to be careful at this point:
\EQ{BBBB} is not necessarily a Bell inequality. It may happen that
$\EE_1 \EE_2$ have correlation terms which contain two or more
variables of the same qubit, e.,g.,
$(X_1Z_2)(Z_1X_2)=X_1Z_1X_2Z_2.$ Such a correlation term cannot
appear in a Bell inequality. Because of that we need the following
theorem.
\\
{\bf Theorem 3.} Let us consider two Bell inequalities
of the form \EQ{tb}. If for each qubit $k$ at most only
one of the inequalities contain variables corresponding
to the qubit, then \EQ{BBBB} describes a composite Bell
inequality.
If both of the inequalities contain variables
corresponding to qubit $k$, then \EQ{BBBB} still describes
a Bell inequality if the inequalities contain the same variable
for qubit $k$.
\\
{\it Proof.}  After the previous discussion it is clear that none of
the correlation terms of the Bell inequality \EQ{BBBB} contain more
than two variables for a qubit. They may contain quadratic terms
such as $X_k^2$, however, these can be replaced by $1.$ $\qed$

Let us now consider two of our Bell inequalities, $\BB(i,I_i)$ and
$\BB(j,I_j),$ where $I_{i/j} \subset \NN{i/j}.$ From now let us
omit the second index after $\BB$. Then based on the previous
ideas a new composite Bell inequality can be constructed with a
Bell operator $\BB:=\BB(i)\BB(j)$ if the qubits in $\{i\}\cup
I_i$ and the qubits in $\{j\}\cup I_j$ are not neighbors
\cite{TG05}. For the composite inequality
$\CC(\BB)=\CC[\BB(i)]\CC[\BB(j)]$ and
$\VV(\BB)=\VV[\BB(i)]\VV[\BB(j)]$ thus the violation of
local realism is larger for the composite inequality than for the
two original inequalities.

Based on these ideas, composite Bell inequalities can be created
from several inequalities. Let us see a concrete example. For an
$n$-qubit cluster state we have the stabilizing operators
$g_i^{(LC_n)}:=Z^{(i-1)}X^{(i)}Z^{(i+1)}$ where $i
\in\{1,2,..,n\}$ and for the boundaries $Z^{(0)}=Z^{(n+1)}=\eins$.
Then we can define the following Bell inequality for vertex $i$
\bea
\BB_i^{(LC_n)}&:=&g_i^{(LC_n)}(1+g_{i+1}^{(LC_n)})(1+g_{i-1}^{(LC_n)})
\nonumber\\
&=&Z^{(i-1)}X^{(i)}Z^{(i+1)}+Z^{(i-2)}Y^{(i-1)}Y^{(i)}Z^{(i+1)}
\nonumber\\
&+&Z^{(i-1)}Y^{(i)}Y^{(i+1)}Z^{(i+2)}
\nonumber\\
&-&Z^{(i-2)}Y^{(i-1)}X^{(i)}Y^{(i+1)}Z^{(i+2)},
\nonumber\\
\VV(\BB_i^{(LC_n)}) &=&2. \eea Now we can combine these Bell
inequalities, for different $i$ as  illustrated in Fig.
\ref{fig_graphs}(b). Here $\BB_2^{(LC_n)}$ and $\BB_6^{(LC_n)}$
are represented by two bold subgraphs. If $n$ is divisible by four
then we obtain a composite inequality characterized by \bea
\BB^{(LC_n)}&:=& \prod_{i=1}^{n/4}\BB_{4i-2}^{(LC_n)},
\nonumber\\
\VV(\BB^{(LC_n)})&=&2^{n/4}. \eea Thus the violation increases
exponentially with $n$.

These ideas can be generalized for a two-dimensional lattice as
shown in Fig.~\ref{fig_graphs}(c). Here $5$-body Bell inequalities,
represented again by bold subgraphs in the figure, can be combined
in order to obtain a violation of local realism increasing
exponentially with the number of vertices. Bell inequalities are
also shown this way for a hexagonal lattice in
Fig.~\ref{fig_graphs}(d). These ideas can straightforwardly be
generalized for arbitrary graphs.

\begin{table}
\caption{Maximal violation $\VV$ of local realism for composite
Bell inequalities for various interesting graph states as a
function of the number of qubits. These composite inequalities are
constructed from the inequalities of Theorem 1.}
\begin{tabular}{|l||c||c|c|c|c|c|c|c|c|c|c|}
\hline Number of qubits& $n$  & 3 & 4 & 5 & 6 & 7 & 8 & 9 & 10 &
11 & 12
\\
\hline \hline Linear cluster graph& ${LC_n}$& 2 &2&
2&2&4&$4$&$4$&$4$&$8$ & $8$
\\
\hline Ring cluster graph & ${RC_n}$& 2 & 2& 2& 2 &2& $ 4$ & $ 4$&
$4$&$4$&$8$
\\
\hline Star graph & ${ST_n}$ & 2 & 2 & 4& 4& 8 & 8 & 16 &16
&$32$&$32$
\\
\hline
\end{tabular}
\end{table}

In Table I the relative violation of local realism is shown for
some interesting graph states as the function of the number of
qubits for Bell inequalities constructed based on the previous
ideas. The state corresponding to a star graph is equivalent to a
GHZ state. The corresponding inequality is equivalent to Mermin's
inequality under relabeling the variables and it has the highest
violation of local realism for a given number of qubits.

\section{Alternative Bell inequalities with multi-qubit variables}

In Theorem 1 we presented Mermin-type Bell inequalities with
multi-qubit observables for graph states. Now we will show that
Ardehali-type inequalities can also be constructed and these have
a higher violation of local realism for odd $|I|$. Note that they
are not constructed from stabilizing terms.
\\
{\bf Theorem 4.} Let us consider vertex $i$ which has
$N_{\rm{neigh}}\ge 2$ neighbors given in $I=\{I_k\}_{k=1}^{N_{\rm{neigh}}}$
such that they are not connected by edges. Let furthermore
$A^{(k)}$ and $B^{(k)}$ be defined as in
Eq.~(\ref{nonlocalmermin}) and define also
\begin{eqnarray}
Q^{(1)}&:=& \frac{A^{(1)}- B^{(1)}}{\sqrt{2}},
\nonumber \\
W^{(1)}&:=& \frac{A^{(1)} + B^{(1)}}{\sqrt{2}}.
\label{QW}
\end{eqnarray}
Then, we can write down the following Bell inequality
\begin{widetext}
\begin{eqnarray}
&&(Q_{1}-W_{1})
\big( - \sum_\pi
A_{2}A_{3}A_{4}A_{5}
\cdot\cdot A_{M}
+ \sum_\pi
B_{2}B_{3}A_{4}A_{5}
\cdot\cdot A_{M}
- \sum_\pi
B_{2}B_{3}B_{4}B_{5}
\cdot\cdot A_{M}
...
\big)
\nonumber
\\
&+&(Q_{1}+W_{1})
\big(\sum_\pi
B_{2}A_{3}A_{4}A_{5}
\cdot\cdot A_{M}
- \sum_\pi
B_{2}B_{3}B_{4}A_{5}
\cdot\cdot A_{M}
+ \sum_\pi
B_{2}B_{3}B_{4}B_{5}
\cdot\cdot A_{M}
...
\big)
\leq {L}_A(|I|+1), \label{Ardehali}
\end{eqnarray}
\end{widetext}
where the bound for the Bell inequality is
\be
{L}_A(m):=
\left\{
\begin{array}{ll}
2^{\frac{m+1}{2}} \,\, & \textrm{ for odd $m$, }
\\
2^{\frac{m}{2}} \,\, & \textrm{ for even $m$. }
\end{array}
\right. \ee Again, $\sum_\pi$ represents the sum of all possible
permutations of the qubits that give distinct terms. If $A_k$,
$B_k$, $Q_{1}$ and $W_{1}$ correspond to the measurement of
quantum operators $A^{(k)}$, $B^{(k)}$, $Q^{(1)}$ and $W^{(1)}$,
respectively, then the graph state $\ket{G}$ maximally violates
Eq.~(\ref{Ardehali}). Note that similarly as in Theorem 1 a party
consists of several qubits here.
\\
{\it Proof.} The bound for LHV models is valid since
Eq.~(\ref{Ardehali}) is the Bell operator of the $(|I|+1)$-body
Ardehali's Bell inequality \cite{ardehali,ARDEHALI} for $Q_1$,
$W_1$, and the $A_k$'s and $B_k$'s. For the same reason, the
maximal value for the Bell operator (\ref{Ardehali}) for quantum
states is the same as for Ardehali's inequality, i.e.,
$2^{|I|}\sqrt{2}.$ This value is obtained for $\ket{G}$ as we will
show. In order to see that let us substitute the definitions of
$Q^{(1)}/W^{(1)}$ given in Eq.~(\ref{QW}) into the Bell operator
of the inequality Eq.~(\ref{Ardehali}). That is, let us substitute
$-\sqrt{2} B^{(1)}$ for $(Q^{(1)}-W^{(1)})$ and $\sqrt{2} A^{(1)}$
for $(Q^{(1)}+W^{(1)})$. Then expand the brackets. This way one
obtains the Bell operator as the sum of $2^{|I|}$ terms. These
terms correspond to stabilizing operators multiplied by
$\sqrt{2}$. $\qed$

\section{Proving that our inequalities are extremal}

Based on Ref.~\cite{WW01}, we know that
our Mermin-type inequalities given in Theorem 1 have a
maximal violation of local realism $\VV$ [defined in Eq.~(\ref{VV})] among $(|I|+1)$-partite Bell inequalities for even $|I|.$ Here this statement is valid for inequalities
analyzed in Ref.~\cite{WW01,ZB02}, which need the measurement of
two observables for each party and they
are the weighted sum of full correlation terms.
The same is true for the
Ardehali-type inequalities given in Theorem 4 for odd $|I|$ \cite{max}. Moreover,
we can prove the following:\\
{\bf Theorem 5.} Our Mermin-type inequalities given in Theorem 1
are extremal for even $|I|$, i.e., they are facets of the convex polytope of
correlations consistent with LHV models. This is also true for
our Ardehali-type inequalities given in Theorem 4 for odd $|I|.$
Combining them one obtains also extremal inequalities.
\\
{\it Proof.} If we did not have many-body observables then it
would be clear that our Bell inequalities are facets \cite{WW01}.
It has also been proved that by multiplying some of these
inequalities with each other, extremal inequalities are obtained
\cite{WW01}. Now, however, we have to prove that by replacing some
of the variables by the products of several variables does not
change this property \cite{lifting}. First of all, one has to
stress that when drawing the convex polytope, the axes correspond
to the expectation values of the many-body correlations terms
appearing in the Bell inequality \cite{polytope}.
Clearly, replacing a variable
with several variables by inserting $Z^{(k)}$'s in some of these
correlation terms does not change the convex polytope of
correlations allowed by local models. The transformed Bell
inequalities also correspond to the same hyperplane as before the
transformation. $\qed$

\section{Comparison with existing Bell inequalities for
four-qubit cluster states}

The systematic study of Bell inequalities for graph states was
initiated in an important paper by Scarani, Ac\'{\i}n, Schenck,
and Aspelmeyer \cite{S04}. Moreover, in this paper,
a Bell inequality for a four-qubit cluster state
was presented which has already been
used for detecting the violation of local realism experimentally \cite{clusterexp}.
The inequality of Ref.~\cite{S04} is also a
Mermin's inequality with composite observables
\begin{equation}
X_1X_3Z_4+Z_1Y_2Y_3Z_4 +X_1Y_3Y_4-Z_1Y_2X_3Y_4 \le 2. \label{S04paper}
\end{equation}
It is instructive to write down its Bell operator with the
stabilizing operator of a cluster state
\begin{equation}
\BB^{(LC_4)}=g_3^{(LC_4)}(g_1^{(LC_4)}+g_2^{(LC_4)})
(\eins+g_4^{(LC_4)}).
\end{equation}
This is different from our ansatz in  Eq.~(\ref{bbi_bell}). The
inequality obtained from our ansatz Eq.~(\ref{bbi_bell}) for $i=3$
is
\begin{equation}
Z_2 X_3 Z_4+Z_1 Y_2 Y_3 Z_4+Z_2 Y_3 Y_4-Z_1 Y_2 X_3 Y_4 \le 2. \label{B2}
\end{equation}\\
The following two four-qubit Bell inequalities are
also built with stabilizing terms and have a factor of two
violation of local realism:
\begin{eqnarray}
X_1X_3Z_4-Y_1X_2Y_3Z_4 +X_1Y_3Y_4+Y_1X_2X_3Y_4\le 2,\nonumber\\
Z_2X_3Z_4-Y_1X_2Y_3Z_4 +Z_2Y_3Y_4+Y_1X_2X_3Y_4 \le 2.\nonumber\\\label{B34}
\end{eqnarray}

Further four inequalities can be obtained by exchanging qubits $1$ and $4$,
and qubits $2$ and $3$, in the previous four Bell inequalities.
These eight inequalities are all maximally violated by
the four-qubit cluster state $\ket{LC_4},$ however, not only by the cluster state.
The maximum of the Bell operator for these inequalities is doubly degenerate.
Thus, as discussed in Ref. \cite{S04} for the case of Eq.~(\ref{S04paper}),
they are maximally violated also by some mixed states.

It can be proved by direct calculation that adding any two of these eight
inequalities another inequality is obtained such
that only the four-qubit cluster state violates it maximally.
Thus from the degree of violation of local realism
one can also obtain fidelity information, i.e., one can get information on how
close the quantum state is to the cluster state.
To be more specific, let us see a concrete example for using this fact.
Let us denote the Bell operators of the four inequalities
in Eqs.~(\ref{S04paper}, \ref{B2}, \ref{B34}) by $\BB_k$ with $k=1,2,3,4.$
Then direct calculation shows that the following matrix is positive semidefinite
\begin{equation}
16\ketbra{LC_4}-\BB_1-\BB_2-\BB_3-\BB_4\ge 0.
\end{equation}
Hence a lower bound on the fidelity can be obtained as
\begin{eqnarray}
F \geq \frac{1}{16}\exs{\BB_1+\BB_2+\BB_3+\BB_4}.
\end{eqnarray}

\section{Conclusion}

We discussed how to construct two-setting Bell inequalities for
detecting the violation of local realism for quantum states close
to graph states. These Bell inequalities allow at least a factor
of two violation of local realism. We used the stabilizer theory
for constructing our inequalities. For several families of states
we have shown that the relative violation increases exponentially
with the size. Some of the inequalities presented are facets of
the convex polytope corresponding to the correlations permitted by
local hidden variable models.

\section{Acknowledgment}

We would like to thank A.~Ac\'{\i}n, J.I.~Cirac, P.~Hyllus  and
C.-Y.~Lu for useful discussions. G.T. especially thanks M.M.~Wolf
for many helpful discussions on Bell inequalities. We also
acknowledge the support of the Austrian Science Foundation (FWF),
the EU projects RESQ, ProSecCo, OLAQUI, SCALA and QUPRODIS, the
DFG and the Kompetenznetzwerk Quanteninformationsverarbeitung der
Bayerischen Staatsregierung. G.T. thanks the Marie Curie
Fellowship of the European Union (Grant No. MEIF-CT-2003-500183)
and the National Research Fund of Hungary  OTKA  under Contract
No. T049234.


\begin{thebibliography}{99}

\bibitem{B64}
J.S. Bell, Physics
{\bf 1}, 195 (1964);
for a review see R. Werner and M. Wolf, Quant. Inf. Comp. {\bf 1 (3)}, 1 (2001);
for results on multipartite Bell inequalities see
Refs.~\cite{mermin,ardehali,PR92,ZB02,WW01,PS01} and G. Svetlichny,
Phys. Rev. D {\bf 35}, 3066 (1987);
A.V. Belinskii and D.N. Klyshko,
Usp. Fiz. Nauk
{\bf 163} (8), 1 (1993);
N. Gisin and H. Bechmann-Pasquinucci, Phys. Lett. A {\bf 246}, 1 (1998);
A. Peres, Found. Phys. {\bf 29}, 589 (1999);
D. Collins, N. Gisin, S. Popescu, D. Roberts, and V. Scarani,
Phys. Rev. Lett. {\bf 88}, 170405 (2002);
W. Laskowski, T. Paterek,
M. \.Zukowski, and {\v C}. Brukner,
Phys. Rev. Lett. {\bf 93}, 200401 (2004).

\bibitem{F81} M. Froissart, Nuovo Cimento B {\bf 64}, 241 (1981).

\bibitem{mermin} N.D. Mermin, Phys. Rev. Lett. {\bf 65}, 1838 (1990).

\bibitem{ardehali} M. Ardehali, Phys. Rev. A {\bf 46}, 5375 (1992).

\bibitem{PR92} S. Popescu and D. Rohrlich, Phys. Lett. A {\bf 166}, 293
(1992).

\bibitem{PS01} I. Pitowsky and K. Svozil, Phys. Rev. A {\bf 64}, 014102 (2001).

\bibitem{ZB02} M. \.Zukowski and {\v C}. Brukner, Phys. Rev. Lett. {\bf 88}, 210401 (2002).

\bibitem{WW01}
R.F. Werner and M.M. Wolf, Phys. Rev. A {\bf 64}, 032112 (2001).

\bibitem{W89} R.F. Werner, Phys. Rev. A {\bf 40}, 4277 (1989).

\bibitem{A02} A. Ac\'{\i}n, private communication;
Phys. Rev. Lett. {\bf 88}, 027901 (2002);
A. Ac\'{\i}n, V. Scarani, and M.M. Wolf, Phys. Rev. A
{\bf 66}, 042323 (2002).

\bibitem{zukbruk} {\v C}. Brukner, M. \.Zukowski,
J.-W. Pan, and A. Zeilinger, Phys. Rev. Lett. {\bf 92}, 127901
(2004).

\bibitem{graph1} W. D\"ur, H. Aschauer, and H.J. Briegel,
Phys. Rev. Lett. {\bf 91}, 107903 (2003).

\bibitem{graph2}
M. Hein, J. Eisert, and  H.J. Briegel,
Phys. Rev. A {\bf 69}, 062311 (2004).

\bibitem{graph3} M. Van den Nest, J. Dehaene, and B. De Moor,
Phys. Rev. A {\bf 72}, 014307 (2005); {\bf 69}, 022316 (2004);
{\bf 70}, 034302 (2004); K.M.R. Audenaert and  M.B. Plenio, New J.
Phys. {\bf 7}, 170 (2005); A. Hamma, R. Ionicioiu, and P. Zanardi,
Phys. Rev. A {\bf 72}, 012324 (2005); D.E. Browne and  T. Rudolph,
Phys. Rev. Lett. {\bf 95}, 010501 (2005).

\bibitem{experiments} For experimental implementations see
Refs.~\cite{graphapp1ex, clusterexp} and
O. Mandel, M. Greiner, A. Widera, T. Rom, T.W. Hänsch and I. Bloch,
Nature (London) {\bf 425}, 937 (2003);
A.-N. Zhang, C.-Y. Lu, X.-Q. Zhou, Y.-A. Chen, Z. Zhao, T. Yang,
and J.-W. Pan, quant-ph/0501036.

\bibitem{GH90} D.M. Greenberger, M.A. Horne,
A. Shimony, and A. Zeilinger, Am. J. Phys. {\bf 58}, 1131 (1990).

\bibitem{cluster}  H.J. Briegel and R. Raussendorf, Phys. Rev. Lett.
{\bf 86}, 910 (2001).

\bibitem{graphapp1th} R. Raussendorf and H.J. Briegel,
Phys. Rev. Lett. {\bf 86}, 5188 (2001); M. Nielsen, {\it ibid.},
{\bf 93}, 040503 (2004).

\bibitem{graphapp1ex} P. Walther, K.J. Resch,
T. Rudolph, E. Schenck, H. Weinfurter, V. Vedral, M. Aspelmeyer,
and A. Zeilinger, Nature (London) {\bf 434}, 169 (2005); N.
Kiesel, C. Schmid, U. Weber, G. T\'oth, O. G\"uhne, R. Ursin, and
H. Weinfurter, Phys. Rev. Lett. {\bf 95}, 210502 (2005).

\bibitem{graphapp2} D. Schlingemann and R.F. Werner,
Phys. Rev. A {\bf 65}, 012308 (2002); M. Grassl, A. Klappenecker,
and M. R{\"o}tteler, in Proc. 2002 IEEE International Symposium on
Information Theory, Lausanne, Switzerland, p. 45; K. Chen, H.-K.
Lo, quant-ph/0404133.

\bibitem{G96} D. Gottesman,  Phys. Rev. A {\bf 54}, 1862 (1996).

\bibitem{S04} V. Scarani, A. Ac\'{\i}n, E. Schenck, and
M. Aspelmeyer, Phys. Rev. A {\bf 71}, 042325 (2005).
\bibitem{DP97} D.P. DiVincenzo and
A. Peres, Phys. Rev. A {\bf 55}, 4089 (1997).
\bibitem{GTHB04}  O. G\"uhne, G. T\'oth, P. Hyllus, and H.J. Briegel,
Phys. Rev. Lett. {\bf 95}, 120405 (2005). 

\bibitem{MERMIN} If one takes the Bell operator
presented in Ref. \cite{mermin} and replaces $A^{(k)}$ and
$B^{(k)}$ by $X^{(k)}$ and $Y^{(k)}$, respectively, then the Bell
operator Eq.~(\ref{BBB}) is obtained.

\bibitem{JIC} J.I. Cirac, private communication.

\bibitem{TG05} In this case the stabilizing terms in $\BB(i)$ and
$\BB(j)$ commute locally. See G. T\'oth and O. G\"uhne,
Phys. Rev. A {\bf 72}, 022340 (2005). 

\bibitem{ARDEHALI} If one takes the Bell inequality
presented in Ref. \cite{ardehali} and interchanges $A_k/B_k$ with
$\sigma_{x/y}^{k-1}$ for $k\ge2$ and $Q_1/W_1$ by
$\sigma_{a/b}^n$, respectively, then inequality
Eq.~(\ref{Ardehali}) is obtained.

\bibitem{max} Ref.~\cite{WW01} in Eq.(25) states that the maximal violation of local realism
for an $n$-partite Bell inequality with full correlation terms is
bounded as $\VV\le 2^{\frac{n-1}{2}}.$ Ref.~\cite{WW01} also
points out that Mermin inequalities for odd number of parties and
Ardehali inequalities for even number of parties give the maximal
$\VV=2^{\frac{n-1}{2}}.$ Note that Ref.~\cite{WW01} uses the term
'the set of inequalities going back to Mermin' in the general
sense, denoting the Mermin and Ardehali inequalities giving
maximal violation mentioned above.

\bibitem{lifting} Stefano Pironio, J. Math. Phys. {\bf 46}, 062112 (2005).

\bibitem{polytope} The coordinate axes could also be the probabilities of the different measurement outcomes rather than the many-body correlation terms.
Such an approach was followed, for example, in Ref.~\cite{PS01}.

\bibitem{clusterexp}
P. Walther, M. Aspelmeyer, K.J. Resch, and A. Zeilinger, Phys.
Rev. Lett. {\bf 95}, 020403 (2005).


\end{thebibliography}
\end{document}